# Two ultra-stable novel allotropes of Tellurium few-layers

Cong Wang[1,†], Linlu Wu[1,†], Xieyu Zhou[1,†], Linwei Zhou[1], Pengjie Guo, Kai Liu[1], Zhong-Yi Lu[1], Zhihai Cheng[1], Yang Chai[2] and Wei Ji[1, *]

[1]*Beijing Key Laboratory of Optoelectronic Functional Materials & Micro-Nano Devices, Department of Physics, Renmin University of China, Beijing 100872, P. R. China*

[2]*Department of Applied Physics, The Hong Kong Polytechnic University, Hung Hom, Kowloon, Hong Kong, P. R. China*

[†]*These authors contribute equally to this work*

\* wji@ruc.edu.cn

At least four two- or quasi-one- dimensional allotropes and a mixture of them were theoretically predicted or experimentally observed for low-dimensional Te, namely the $\alpha$, $\beta$, $\gamma$, $\delta$ and chiral-$\alpha$+$\delta$ phases. Among them the $\gamma$ and $\alpha$ phases were found the most stable phases for monolayer and thicker layers, respectively. Here, we found two novel low-dimensional phases, namely the $\varepsilon$ and $\zeta$ phases. The $\zeta$ phase is over 29 meV/Te more stable than and the $\varepsilon$ phase shows comparable stability with the most stable monolayer $\gamma$ phase. The energetic difference between the $\zeta$ and $\alpha$ phases reduces with respect to the increased layer thickness and vanishes at the four-layer (12-sublayer) thickness, while this thickness increases under change doping. Both $\varepsilon$ and $\zeta$ phases are metallic chains and layers, respectively. The $\zeta$ phase, with very strong interlayer coupling, shows quantum well states in its layer-dependent bandstructures. These results provide significantly insight into the understanding of polytypism in Te few-layers and may boost tremendous studies on properties of various few-layer phases.

# Introduction

Low dimensional elemental materials are a large family of two-dimensional materials[1-3]. Graphene[4-8] was the first mono-layer ever isolated for carbon and 2D materials while graphdiyne[9] is an allotrope of it. Layers of group IV elements, known as silicene[10-12], germanene[13] and stanene[14, 15], and layers of a group III element, i.e. borophenes[16-18], as well as group V few-layers, i.e. 2D P[19-23], As[24], Sb[25] and Bi[26], were subsequently predicted and synthesized or isolated. These mono- and few-layers of groups III, IV and V elements were experimentally shown or were theoretically predicted to have tunable bandgap[7, 12, 21, 24], high carrier mobility[5, 19, 20, 23, 25], strong light absorption and response in infrared and visible lights ranges[27, 28], exceptional mechanical and frictional properties[5, 17, 22], catalysis activities[29], topological electronic states[8, 10, 14, 26], superconductivity[30, 31] and among the others[32]. However, the few-layer forms of group VI elements are still ambiguous and are yet to be unveiled.

Tellurium few-layers are a category of emerging group VI 2D layers[33-40]. The few-layer α phase, cleavable from its bulk counterpart, show amazing electronic, optical and vibrational properties and can be synthesized using wet-chemistry methods[35]. Therefore, the synthesis of it does not need a substrate while other elemental few-layers needs either exfoliation from their bulk counterparts [4-6,20, 23, 41-44], or substrates to stabilize[45, 46] or assist[17] the synthesis of the layers. A striking feature of it lies in that it has, at least, four few-layer allotropes predicted by DFT calculations, the number of which is comparable with that of carbon. A previous theory shows that meta-stable few-layer phases could be stabilized with charge doping[47]. Another theory predicted topological states in Te nanostructures[48], however, those structures are highly unstable. It would be thus of interest to know if there are any new phases, preferably with topological states, yet to be unveiled and the reason why Te could offer so many allotropes. Answers to these questions would boost both fundamental research and device applications.

Here, we predicted two novel forms, i.e. $\varepsilon$ and $\zeta$ phases, of Te few-layers, among which the $\zeta$ phase shows extraordinary stability that its monolayer is 29 meV/Te more stable than the most-stable $\gamma$ monolayer and its bilayer is over 30 meV/Te more stable than the most-stable $\alpha$ bilayer. An energetic crossover between the $\zeta$ and $\alpha$ phases occurs at the four-layer (12-sublayers) thickness that the $\zeta$ phase is prone to transform into the $\alpha$ phase beyond that thickness, while either hole or electron doping stabilizes the $\zeta$ phase and pushes the crossover to thicker layers. The $\varepsilon$ phase is less stable than the $\zeta$ phase, but has a comparable stability with the monolayer $\gamma$ or few-layer $\alpha$ phase. Phonon dispersion calculations suggest that the free-standing forms both phases are stable and could be exfoliated from thicker layers or substrates. These two novel phases strongly promote subsequent studies on polytypism of Te few-layers and add two new members to the family of Te allotropes.

**Results and discussion**

The $\alpha$-phase, comprised of helical chains bonded with CLQB along inter-chain directions, is the most stable form in Te few-layers and bulk [34, 40] (see Supplementary Fig. S1a). In monolayers, however, the $\gamma$-phase (Fig. 1a) was believed the most stable phase and the $\alpha$-phase becomes unstable and spontaneously transforms into the $\beta$-phase (Fig. 1b). The $\gamma$-phase contains rhomboid chains along two directions forming a network in the $C_{3v}$ symmetry while the $\beta$ monolayer is comprised of parallel rhomboid chains with an inclination angle of 29.3° to the *xy*-plane. Strong charge doping increases the inclination angle of the rhomboid chains in the $\beta$-monolayer and the angle eventually reaches 90° at a doping level of 0.50 *e*/Te, giving rise to a new phase $\varepsilon$ (Fig. 1c). This phase has parallel aligned diamond chains, in which Te-Te distance is 3.04 Å and angles are 87° and 93°, respectively, while the interlayer CLQB length is 3.59 Å. Although the $\varepsilon$-monolayer is 6 meV/Te less stable than the $\gamma$-monolayer (Fig. 1e), the relative stability reverses (23 meV/Å$^2$) if comparing surface energies of these two phases, ascribed to the much smaller surface area of the ε-monolayer resulted from its perpendicularly tilted rhomboid chains.

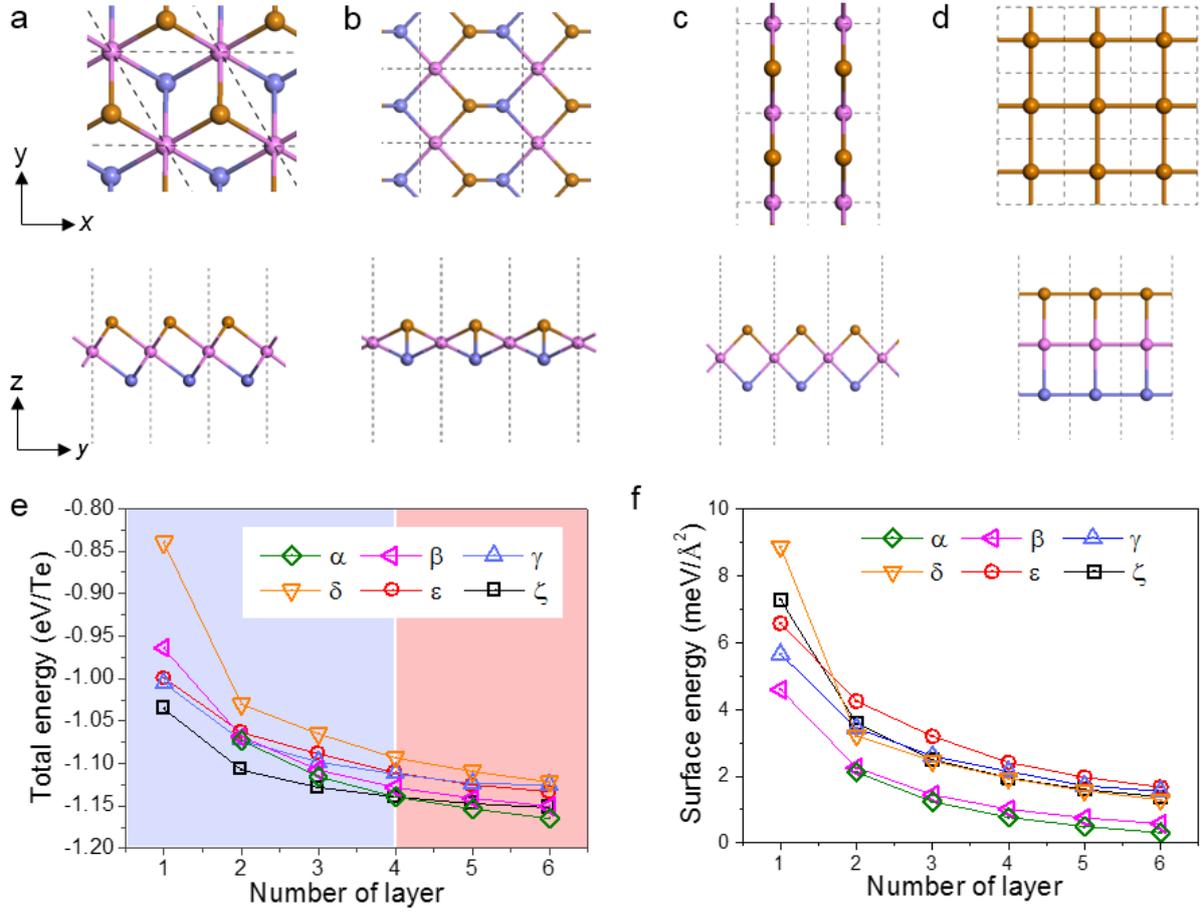

**Figure 1.** (a-d). Top- and side- views of monolayer Te in $\gamma$, $\beta$, $\varepsilon$ and $\zeta$ phase, respectively. Orange, violet and slate-blue balls represent Te atoms in different sublayers along the interlayer direction $z$. (c-d) Total energies per Te atom and surface energies per unit area in different phases as a function of number of sublayers, respectively. Te in $\alpha$, $\gamma$ and $\zeta$ phase are presented with green, slat-blue and black symbols, respectively.

An even more stable $\zeta$ phase (Fig. 1d) was found by relaxing atomic coordinates from laterally shifted $\varepsilon$ layers. The $\zeta$ monolayer consists of three sublayers where the Te atoms form a square lattice, with Te-Te bond length of 3.15 Å, in each sublayer, leading to a structure with an ultra-low specific area and a high symmetry (P4/MMM). A similar, but strongly tilted, structure was previously found in bulk Te under a high pressure of over 8 GPa[49], which transforms into the $\zeta$ bulk phase after the pressure is fast released (Supplementary Fig. S2). Although the $\zeta$ bulk form is less stable than the $\alpha$ bulk phase, the $\zeta$ monolayer and bilayer are much stable with at least 29 meV/Te (93 meV/Å$^2$) and 35 meV/Te (200 meV/Å$^2$) energy gains from the previously believed most-stable γ monolayer and α bilayer, respectively. It would be thus interesting to find the transition boundary of energetic stability between the α and γ multilayers.

The ζ few-layers prefer an AA stacking by at least 9 meV/Te, in which a Te atom of an upper sublayer sits right over another Te atom underneath (see Supplementary Fig. S3 and Table S1). We thus adopted the AA stacking in following calculations. Figure 1e plots the total energies of the six known phases as a function of the number of sublayers, which shows that the ζ few-layer (blue square) is energetically more stable than other five phases before the thickness reaches 12 sublayers (four layers). Beyond this thickness, the structure of the ζ phase still holds but the α phase becomes the most stable phase; this is, most likely, ascribed to weakened surface effects as the bulk properties dominate the behavior of the ζ phase in thicker layers. We also plotted surface energies in Fig. 1f. It shows the α phase is the easiest one to cleave and the β phase has a comparable surface energy. Other phases, except the ε phase, show slightly higher but reasonable surface energies.

The intra-layer bond lengths (lattice constant *a/b*) are 3.02 Å and 3.08 Å in a mono- and bi-atomically thick ζ sublayers, respectively, which are much smaller that the bulk value of 3.21 Å. Figure 2a shows the evolution of intra-sublayer and inter-sublayer bond lengths as a function of layer thickness. The increased thickness significantly varies both the inter-sublayer and the intra-sublayer distances, indicating a strong inter-sublayer interaction. The more the sublayers stacked together, the stronger the charge transfers from $p_z$ orbitals of Te atoms to in-plane $p_x/p_y$ orbitals and intra-sublayer regions, leading to undercut intra-sublayer and reinforced inter-sublayer Te-Te bonds. The intra-sublayer lattice constant, as a result, expands 3.02 Å (1-sublayer) to the bulk value of 3.21 Å at 12-sublayer while the inter-sublayer distance shrinks from 3.38 Å (2-sublayer) to 3.21 Å also at 12-sublayer. Both intra- and inter-sublayer bond lengths converge to the bulk values (3.21 Å) at 12 sublayer, consistent with energetic crossover and the order of stability of the α and ζ bulk forms.

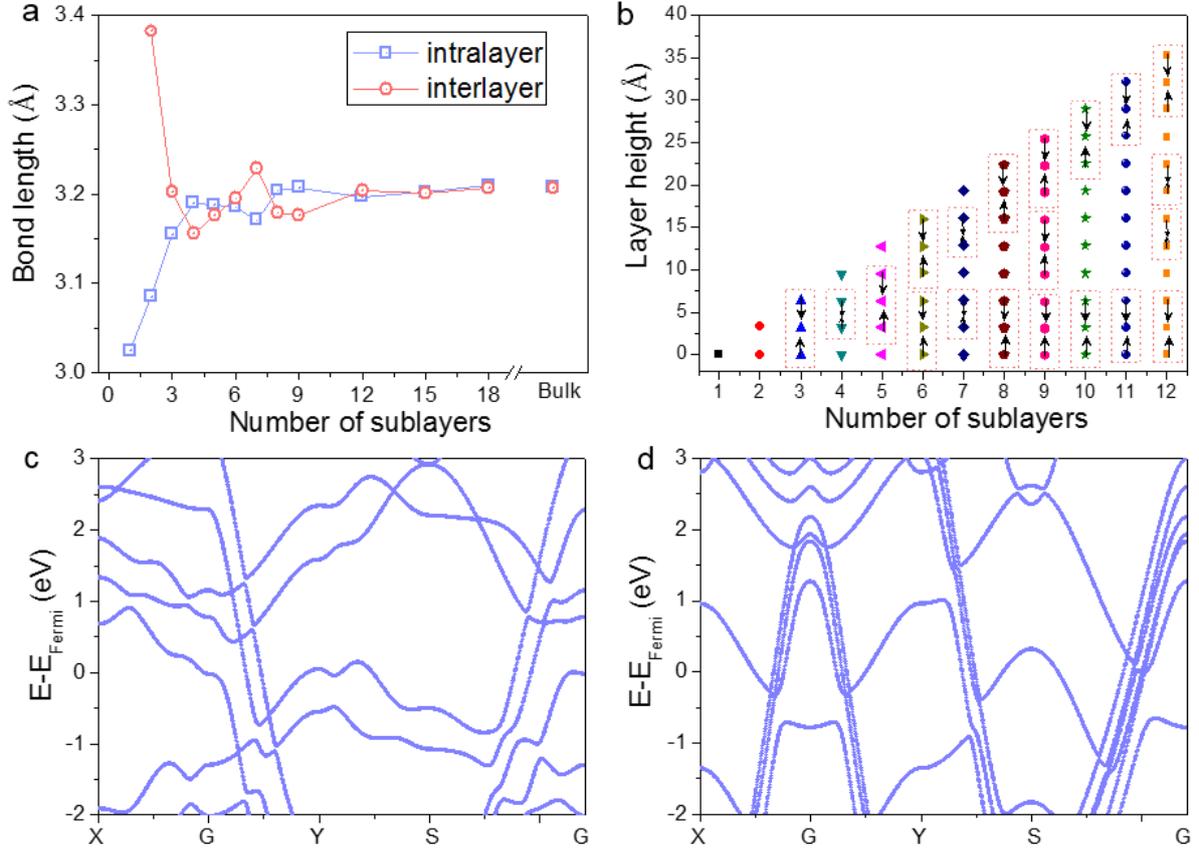

**Figure2. Structure evolution of ζ Te after layer stacking.** (a) Bond lengths as a function of the number of sublayers. The blue and red lines correspond to intra- and average interlayer Te-Te bond lengths, respectively. (b) The evolution of layer heights in ζ few-layer with respect to sublayer number. The layers marked in red dotted rectangular frame tend to form a dimer or trimer when stacking together. (c,d) Electronic band structures of mono-layer ε and ζ calculated with the PBE functional including spin-orbit coupling, respectively. The Fermi energy is set to zero.

It is exceptional that structural relaxations were found in ζ multilayers that they are prone to form dimers or trimers with adjacent sublayers along the interlayer $z$ direction. We used the bulk bond length of 3.21 Å as a criterion. Sublayers with bond lengths smaller than this value were regarded dimerized and trimerized. Figure 2b presented the detailed distributions of dimers or trimers from mono- to 12-sublayers. Sublayers dimerizing or trimerizing together are marked by red dotted rectangles and the directions of atomic relaxations are indicated with black arrows. A trimer first appears in the tri-sublayer (a ζ monolayer), which is, most likely, due to a Fermi surface nesting induced electronic structure and geometry instability. Dimers, trimers and their mixtures emerge in thicker ζ sublayers with the thickness up to 12 sublayers. We tested different combinations of the dimers and trimers confirming the configurations shown in Fig. 2b are the most stable ones (see Supplementary Fig. S4). All of them show mirror and central inversion symmetries along the inter-sublayer direction. Besides that, dimers would not show up at the surface region, which is consistent with the non-dimeric 2-sublayer ζ. The reason why these relaxations occurs is another research topic than we will discuss it elsewhere.

Figures 2c and 2d show the electronic bandstructures of the ε and ζ phases, respectively, where several linear band inversions and their induced bandgaps were observed. Details SOC induced bandgap opening and the inversions were available in Supplementary Figs. S5 and S6. However, these phases were found topologically trivial, as summarized in Supplementary Table S2-S3. Quantum well states were explicitly observed for the states along the z direction, the direction normal to the layer planes, as shown in Supplementary Figs. S7 where shows the evolution of the band structures of ζ few-layers with different thickness.

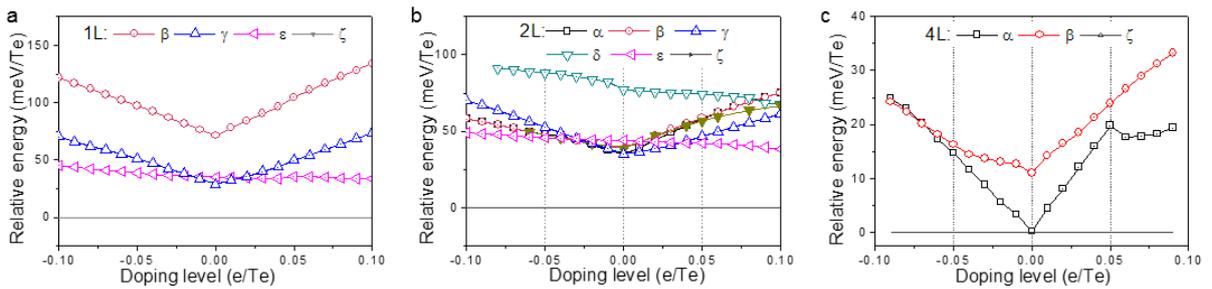

**Figure 3. Phase diagram of Te under charge doping.** Relative total energy of mono- (a), bi- (b) and four-layer (c) Te in different phases as a function of electron/hole doping level. The total energies of the ζ phase were chosen as the energy reference. Lines with different colors correspond to the relative energies of different phases: $\alpha$, black; $\beta$, red; $\gamma$, blue; $\delta$, dark cyan; $\varepsilon$, magenta; $\zeta$, dark gray.

In a recent work[47], we found direct charge doping may change the relative stability of Te phases and may transform a certain phase to another. We thus considered the stability of both the ε and ζ phases under electron or hole doping. Figure 3a shows the total energies of these two phases, γ and β phases in Te monolayers, in which the most stable ζ phase was chosen as the reference zero. The diagram shows that the ζ phase keeps its exceptional stability, with at least 30meV/Te to the β phase, under either electron or hole doping. This statement was double confirmed by the SCAN+rVV10 functional calculations (Supplementary Fig. S8). Further calculations show the ζ monolayer should be the most stable phase even when the doping level reaches 0.8 e/Te or 0.8 h/Te. Such values are beyond the capability of modern ionic liquid gating techniques, which, together with vibrational frequency calculations, guarantee the superior stability of the ζ monolayer. These results indicate that the ζ monolayer has little chance transforming into other forms of low-dimensional Te if it is fabricated. The ε monolayer is slightly less stable than the γ phase at the neutral state, while it becomes more stable when adding or removing electrons, which is substantially different from the previous found δ and mixed phases[47]. This structural phase transition manipulates the semiconducting β or γ phase transforming into the metallic ε phase. Details discussion for the electronic properties of ε phase and the origin of β-ε phase transition in the Te monolayer are available in Supplementary Fig. S9. Figure 3b shows the relative energies of different phases in 2L. It, again, shows the superior stability of the ζ phase and the more stable ε phase than the γ phase under doping. These results suggest doping could further stabilize the ζ phase, which pushes the ζ-α crossover thickness beyond 4L under charge doping as shown in Figure 3c.

**Conclusion**

In summary, we predicted two new low-dimensional Te allotropes, i.e. ε and ζ, which, especially the ζ phase, yield extraordinary stability. It has strong vertical inter-sub-layer interaction that shows quantum well states along the direction normal to the layers. As the most stable few-layer phase found so far, it was surprising that this phase has not been synthesized yet; this is, most likely, due to its substantially different geometry from the helical bulk-like form or the lack of a square substrate lattice. We expected that the ζ phase might be prepared by molecular beam epitaxy, physical vapor deposition, laser or electron beam deposition or even chemical vapor deposition with precisely controlled dosing rates, temperatures, substrates or from a fast released high pressure phase. Unlike semiconducting *α*, *β* and *γ* layers and the *δ* chain, the ζ phase is a metallic layer with high and tunable density of state and strong band dispersion, which is ideal for applications of layered electrodes. Our results added two more allotropes to few-layer Te and open a new avenue for studying topological properties in group VI 2D layers.

**Methods**

Density functional theory calculations were performed using the generalized gradient approximation for the exchange-correlation potential, the projector augmented wave method [50, 51] and a plane-wave basis set as implemented in the Vienna *ab-initio* simulation package (VASP) [52] and Quantum Espresso (QE) [53]. Density functional perturbation theory was employed to calculate phonon-related properties, including Raman intensity (QE), activity (QE) and shifts (VASP), vibrational frequencies at the Gamma point (VASP) and other vibration related properties (VASP). The kinetic energy cut-off for the plane-wave basis set was set to 700 eV for geometric and vibrational properties and 300 eV for electronic structures calculation. A *k*-mesh of 15×11×1 was adopted to sample the first Brillouin zone of the conventional unit cell of few-layer Te in all calculations. The mesh density of *k* points was kept fixed when calculating the properties for bulk Te. In optimizing the system geometry and vibration calculations, van der Waals interactions were considered at the vdW-DF [54, 55] level with the optB88 exchange functional (optB88-vdW) [56-58], which was proved to be accurate in describing the structural properties of layered materials [23, 59-61]. The shape and volume of each supercell were fully optimized and all atoms in the supercell were allowed to relax until the residual force per atom was less than $1\times10^{-4}$ eV·Å$^{-1}$. Electronic bandstructures were calculated using PBE functional and hybrid functional (HSE06) [62, 63] with and without spin-orbit coupling (SOC).

Charge doping on Te atoms was realized with the ionic potential method [64], which was used to model the chare transfer from graphite substrates. For electron doping, electrons are removed from a 4*d* core level of Te and placed into the lowest unoccupied band. For hole doping, electrons were removed from the valence band by adding a negative potential into the 4*d* core level of those three Te atoms. This method ensures the doped charges being located around Te atoms. It also keeps the neutrality of the whole supercell without introducing background charge, which eliminates effects of compensating charges.

## Conflict of interest
The authors declare no conflict of interests.

## Acknowledgments
This project was supported by the National Natural Science Foundation of China (Grant Nos. 11274380, 91433103, 11622437, 61674171 and 61761166009), the Fundamental Research Funds for the Central Universities of China and the Research Funds of Renmin University of China (Grant No. 16XNLQ01), the Research Grant Council of Hong Kong (Grant No. N_PolyU540/17), and the Hong Kong Polytechnic University (Grant Nos. G-SB53). C.W. was supported by the Outstanding Innovative Talents Cultivation Funded Programs 2017 of Renmin University of China. Calculations were performed at the Physics Lab of High-Performance Computing of Renmin University of China and the Shanghai Supercomputer Center.

## Appendix. Supplementary materials
Supplementary materials accompanying this article at.

# Supporting Information

# Two ultra-stable novel allotropes of Tellurium few-layers

Cong Wang[1,†], Linlu Wu[1,†], Xieyu Zhou[1,†], Linwei Zhou[1], Pengjie Guo, Kai Liu[1], Zhong-Yi Lu[1], Zhihai Cheng[1], Yang Chai[2] and Wei Ji[1, *]

[1]*Beijing Key Laboratory of Optoelectronic Functional Materials & Micro-Nano Devices, Department of Physics, Renmin University of China, Beijing 100872, P. R. China*

[2]*Department of Applied Physics, The Hong Kong Polytechnic University, Hung Hom, Kowloon, Hong Kong, P. R. China*

[†]*These authors contribute equally to this work*

\* <wji@ruc.edu.cn>

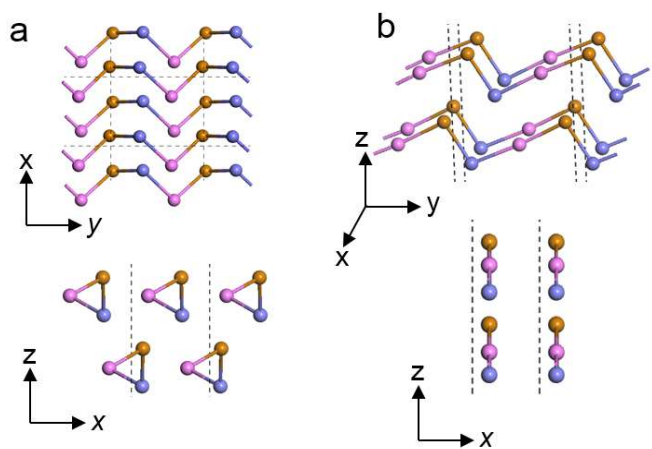

**Supplementary Figure S1. Atomic models of α and δ bilayers.** Figure S1a shows the top- and side-view of the α bilayer. The α phase is comprised of helical chains bonded with CLQB along both inter- and intra-layer directions. The δ phase is comprised of zigzag chains parallel with each other, as shown in Figure 1b. The δ phase can be introduced into mono- and few-layer by uniaxial strain or electron doping.

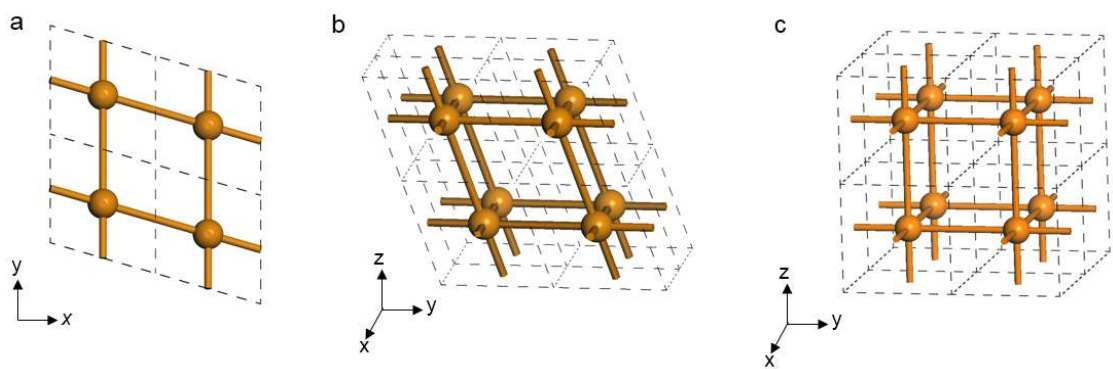

**Supplementary Figure S2. Atomic models of Te bulk under high pressure.** Figure S2a and S2b show the top- and perspective-view of Te bulk under a high pressure of 11.5 GPa (Te-HP). Te-HP takes a rhomboheral lattice with a lattice constant of 2.95 Å and an angle of 102.69°. After the pressure is fast released, the Te-HP spontaneously transforms into ζ phase (cubic lattice) with an elongated lattice constant of 3.21 Å and an angle of 90° (Figure S2c).

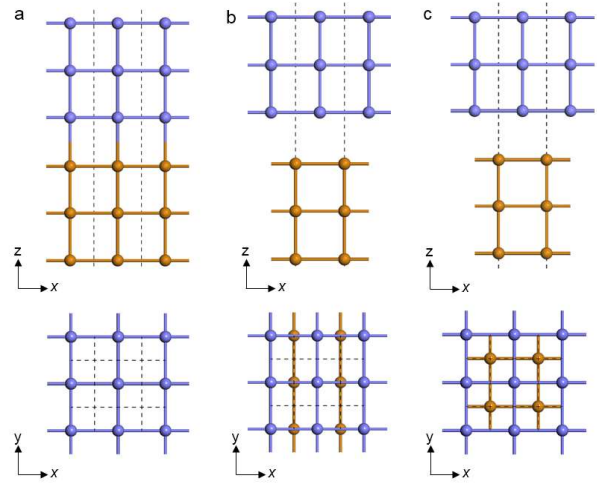

**Supplementary Figure S3. Schematic models of ζ bilayer with stacking orders AA (a), bridge (b) and hollow (c).** Exact data of relative energy can be found in Supplementary Table S1.

**Table S2.** Relative total energies $\Delta E$ (with respect to the most stable configuration) calculated with functionals b88+vdW and SCAN+rVV10. AA stacking shown in Figure S3a was found to be more stable than bridge and hollow stacking with both functionals.

| $\Delta E$ (meV/Te) | b88 +vdW | SCAN +rvv10 |
|---|---|---|
| ζ-bridge | 11 | 12 |
| ζ-AA | 0 | 0 |
| ζ-hollow | 9 | 10 |
| alpha | 35 | 22 |
| beta | 39 | 25 |
| epsilon | 44 | 56 |

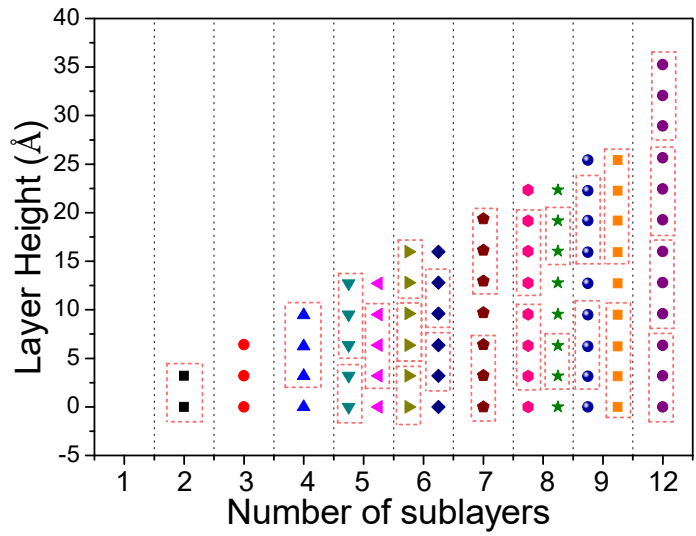

**Supplementary Figure S4. Different combinations of dimers and trimers in ζ few-layer.** Firstly, we set the initial configurations with same a interlayer distance of 3.21 Å and fully optimized the structures. Dimers and trimers showed up after relaxations and the relaxed structures were indeed the most stable configurations shown in Fig. 2c. To further verify the ground states in fewlayer ζ, we tested various configurations with different combinations of the dimers and trimers from bi- to 12-sublayers. Take tri-layer for example, we set the intra-sublayer lattice constant as 3.38 Å, which is the one from bilayer, and found it prone to form trimer after relaxation. Other combinations proposed all showed similar results and transformed to the most stable one. Detailed information about the checked configurations can be seen in Figure S4.

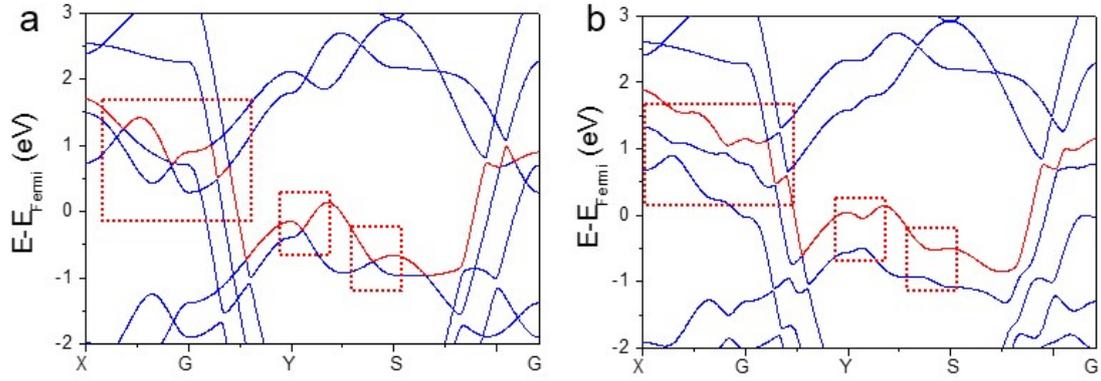

**Supplementary Figure S5. Electronic structures of few-layer ε.** Figure S9a and S9b show band structures of monolayer ε without and with SOC, respectively. Electron doping can induce the semiconducting β phase transforming into the metallic ε phase. Gap opening can also be observed in epsilon phase and Z2 topological invariant calculations reveal it also with trivial topological properties (see Supplementary Table S2). .

**Table S2.** Parities of filled states of monolayer ε at four time-reversal-invariant points in the Brillouin zone. The "+" and "-" correspond to even and odd parity, respectively. Number 1 to 17 represent the number of occupied bands.

| No. | S | Y | G | X | Total |
|---|---|---|---|---|---|
| 1 | + | - | + | - | + |
| 3 | - | + | - | + | + |
| 5 | + | - | + | - | + |
| 7 | - | + | - | + | + |
| 9 | - | - | - | + | - |
| 11 | + | + | + | - | - |
| 13 | - | + | + | - | + |
| 15 | - | + | - | + | + |
| 17 | + | - | - | + | + |
| 19 | + | - | - | + | + |
| Total | + | - | - | + | (+) |

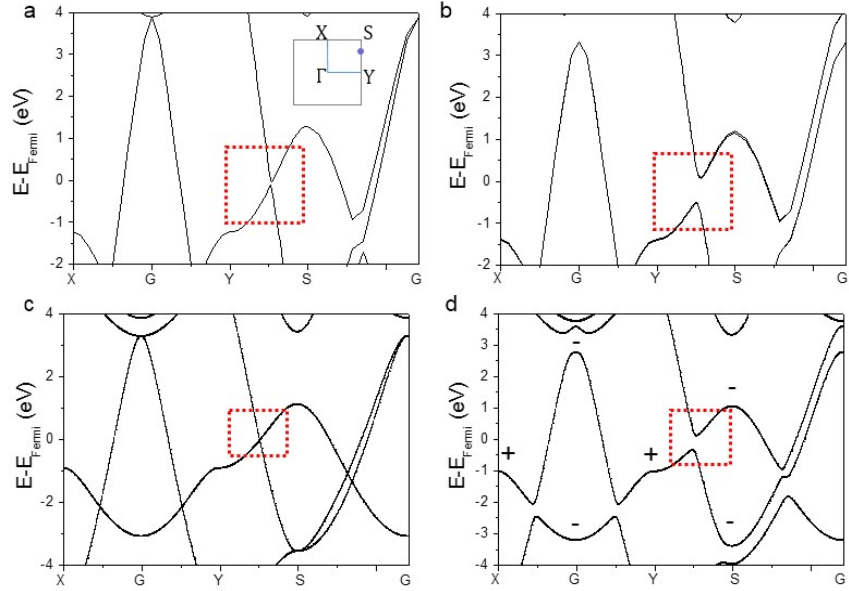

**Supplementary Figure S6. Topological properties of one atom thick ζ Te.** (a-b) Band structures of the single atom layer ζ calculated by HSE06 without (a) and with (b) spin-orbit coupling (SOC), respectively. Red dashed rectangular marks the location of the band inversion and band opening with the SOC effect. Brillouin zone of 2D ζ is shown in the inset figure in **a** and four time reversal invariant points are labeled as Γ, X, Y and S. The blue dot between Y and S represent the point Y' (0.26, 0.50, 0) where gap opening occurs. (c) Band structure of 1-sublayer ζ calculated by PBE including the SOC effect. The band structure of the trimerized ζ tri-sublayer (the ζ monolayer) suggests likely topological electronic properties. We thus examined the band structure of the simplest ζ mono-sublayer to rule out any influence of interlayer interactions. We found several band inversions as confirmed by the orbital decomposed band structures. Among these inversions, the one occurred at the Y' point (0.26, 0.50, 0.00) between Y and S is of particular interest that the inversion point of those two bands sits roughly at the Fermi level. Inclusion of SOC opens a bandgap of 0.68 eV around the Fermi level at the Y' point. We thus calculated the $Z_2$ topological invariant using Quantum Espresso (QE) to verify the topological characteristic of the ζ mono-sublayer. The Te atom in the ζ mono-sublayer is in a square network structure, which has both time reversal and space central inversion symmetries. Therefore, the $Z_2$ topological invariant can be obtained by multiplying parities of filled states at all time-reversal invariant points. Our calculations revealed the $Z_2$ value of 1, which indicates the ζ mono-sublayer with trivial characteristic (see Supplementary Table S2).

**Table S3** Parities of filled states of 1-sublayer ζ at four time-reversal-invariant points in the Brillouin zone. The "+" and "-" correspond to even and odd parity, respectively. Number 1 to 5 represent the number of occupied bands.

| No. | X | Γ | Y | S | Total |
|---|---|---|---|---|---|
| 1 | - | + | - | + | + |
| 3 | + | - | + | - | + |
| 5 | + | - | + | - | + |
| Total | - | + | - | + | (+) |

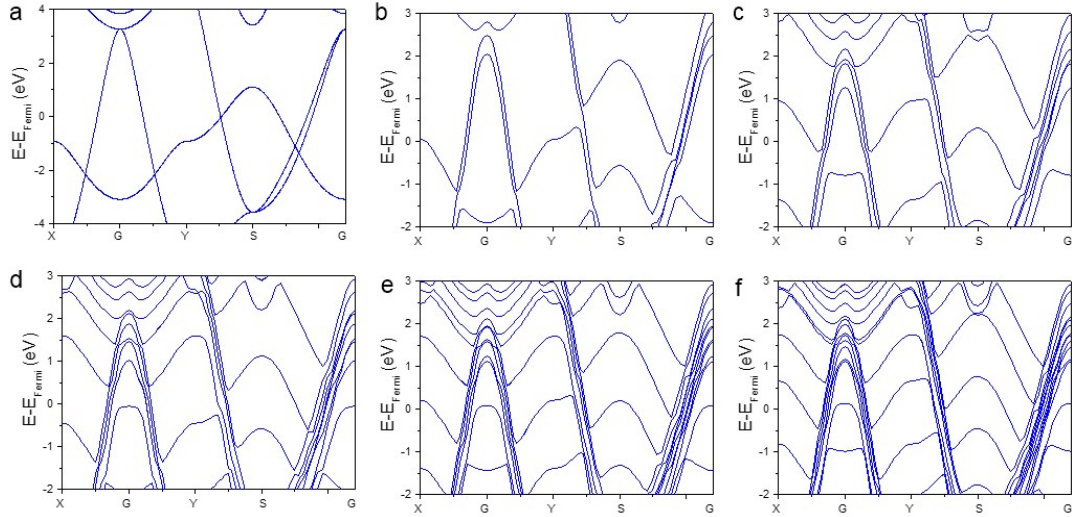

**Supplementary Figure S7. The evolution of the band structures of ζ few-layers with different thickness.** The band structures of ζ from mono- to 6-sublayers calculated by PBE including the SOC effect are shown in a-f, respectively. The nodal lines beside G and S can be observed with an in increased number with sublayer stacking.

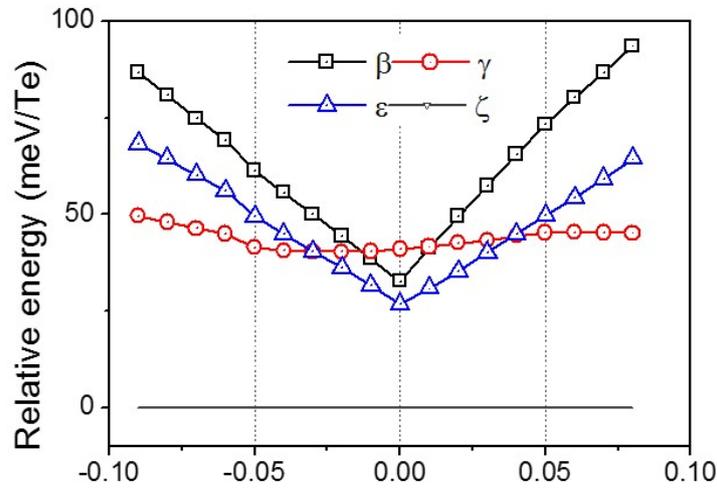

**Supplementary Figure S8. Relative total energies of monolayer Te in different phases as a function of electron/hole doping level calculated with SCAN+rVV10.** The total energies of the β-bilayer were chosen as the energy reference. ζ (blue) is also proved to be stable under charge doping with meta-GGA function of SCAN+rVV10.

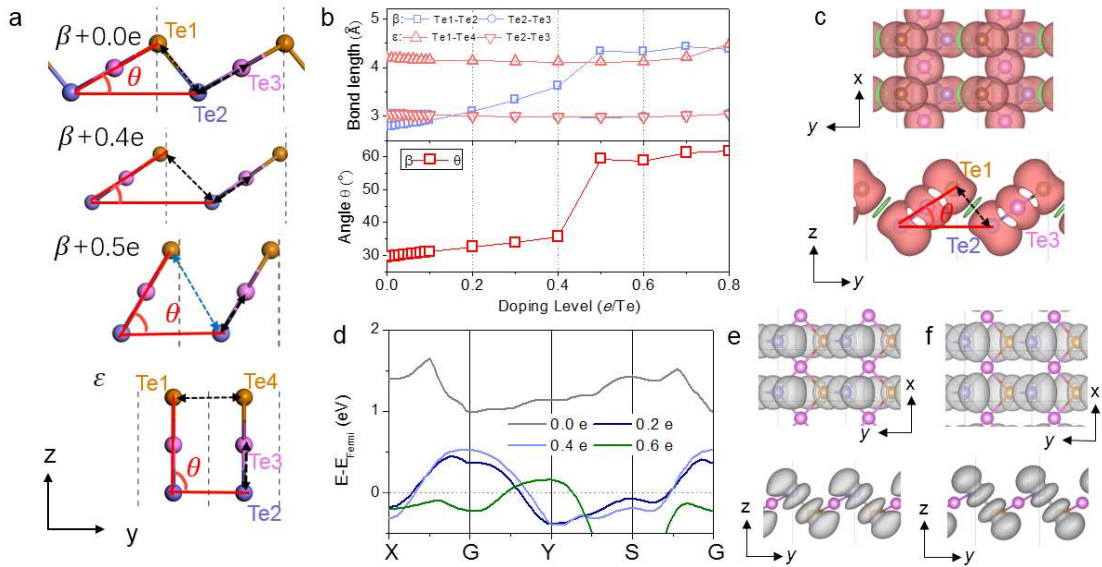

**Supplementary Figure S9. Details of the β-ε phase transition in a Te monolayer.** Under an electron doping level of 0.5e/Te, a β-ε phase transition can be introduced into monolayer Te. β monolayer is comprised of monolayer is comprised of parallel rhomboid chains with an inclination angle of 29.3° to the xy-plane. As shown in Figure S9a and S9b, the inclination angle (red cubics) and inter-chain distance $Te_1$-$Te_2$ (blue cubics) get bigger with respect to doping level and a sudden increase from 35° to 60° can be observed at 0.5e/Te. The intra-chain bond length $Te_2$-$Te_3$ (blue circles) keep almost unchanged under electron doping, indicating that electron doping mainly manipulates the inter-chain coupling instead of change the intra-chain structure. The electron doped β is unstable according to frequency calculation and would transform into ε phase, where the parallel rhomboid chains are perpendicular to the xy-plane. The ε monolayer would keep stable when doping or withdrawing electrons and bond lengths change little under electron doping. Figure S9c plots an electron-doped differential charge densities (DCD) plotted between a 0.4 e/Te doped and a neutral β-bilayers. Charge reduction was found at inter-chain region between $Te_1$ and $Te_2$, which indicates the weakened the $Te_1$-$Te_2$ bonds and rotated rhomboid chains. Such charge redistribution could be explained by doping-dependent band structures shown in Figure S9d. At doping levels under 0.4 e/Te, doped electron mainly filled the state at X and Y. We depicted the wave function norm of the CB1 state at X and Y in Figure S9e and S9f, which are both anti-bonding state of $Te_1$-$Te_2$. Doping electron on these states thus result in the weakening of inter-chain $Te_1$-$Te_2$ bonds and enhance the repulsion between chains, , thus leading to the phase transition.